\documentclass[12pt,draftclsnofoot,onecolumn]{IEEEtran}
\IEEEoverridecommandlockouts 
\ifCLASSINFOpdf
\usepackage[pdftex]{graphicx}
\graphicspath{{./Images/}}
\DeclareGraphicsExtensions{.pdf,.jpeg,.png}
\else
\fi
\usepackage[cmex10]{amsmath}
\usepackage{siunitx}
\usepackage{cite}
\usepackage{psfrag}
\usepackage[utf8]{inputenc}
\usepackage[T1]{fontenc}
\usepackage{amsmath,amsfonts,amsbsy,amssymb}
\usepackage{mathabx}
\usepackage{mathrsfs}
\usepackage[nolist]{acronym}
\usepackage{tabularx}
\usepackage{amssymb}
\usepackage{amsmath}
\usepackage{graphicx}
\usepackage{cite}
\usepackage{multirow}
\usepackage{wasysym}
\usepackage{multirow}
\usepackage{float}
\usepackage{color}
\usepackage{tikz,lipsum,lmodern}
\usepackage[most]{tcolorbox}
\usepackage{hyperref}

\begin{document}

\title{\huge Multi-layer Analysis of IoT-based Systems}
\author{Pedro H. J. Nardelli and Florian Kühnlenz
\thanks{The authors are with the Centre for Wireless Communications (CWC) at University of Oulu, Finland. This work is partly funded by Finnish Academy  by Strategic Research Council/Aka BCDC Energy (n.292854). This material first appears in the \textit{14th International Symposium on Wireless Communication Systems} (ISWCS), Bologna, Italy, 2017. Contact: pedro.nardelli@oulu.fi}%
}

\maketitle

\begin{abstract}
Information and communication technologies are widespread now; it is difficult to exemplify one aspect of our daily life that is completely unrelated to the internet. In this context, Internet of Things is an ongoing step towards the automatization of human activities, including not only information interchanges but also material activities and decision-making processes. Traditional approaches in technological and other sciences often reduce the problems, which may hide emergent features of the system in its totality. One example is frequency control in electricity systems by smart appliances. As a policy, autonomous IoT-based home appliances may decide to postpone or advance their cycles, following the same signal (frequency). If the decisions are selfish (i.e. without considering that other devices may act similarly), the strategy designed to stabilize the system turns out to have the opposite effect. Therefore, a different approach is needed to capture features that emerge from interactions of elements based on information.

This document provides a theoretical-methodological ground to sustain the idea that the IoT builds the structure of awareness of large-scale infrastructures viewed as techno-social cyber-physical systems, which are  special cases of self-developing reflexive-active systems.
As the last phrase already indicates, we need to go through a series of explanations before reaching the point of being capable of analyzing the dynamics of IoT-based systems, constituted by physical, information and regulatory layers.
We expect through this text to clarify what is the structure of awareness  by revisiting the little known Lefebvre's notation. 
From this standpoint, we can analytically show systemic differences that appears when agents using information about the physical system and/or about the other agents (re)act within the system itself, determining then the actually realized system dynamics.
We provide an example of how to carry out this kind of research using the example of smart appliances as a form of stabilizing the grid frequency.
\end{abstract}

\newpage
	
\section{Introduction}
\label{sec:poker}
Forget about IoT, let us start with Poker.
Wikipedia gives following description \cite{poker}:
\begin{quote}
\textit{Poker is a family of card games that combine gambling, strategy, and skill. All poker variants involve betting as an intrinsic part of play, and determine the winner of each hand according to the combinations of players' cards, at least some of which remain hidden until the end of the hand.}
\end{quote}
Individual (good) players are  strategic: they are always processing information from the game table to guide their actions to become the final winner (i.e. the last one in the table).
As any card game, there are prescribed rules and an intrinsic aleatory component in the system dynamics.
Looking as an observer, the matches' dynamics come from the interactions between individual players inside the game.

\vspace{2ex}
\begin{tcolorbox}[colback=black!5!white,colframe=black!50!black,  colbacktitle=black!75!black,title= To think about]
  \begin{itemize}
   \item What would happen if two players are able to communicate during the game?
   \item What happens when players know the betting strategies of others?
   \item Why some players bluff every now and then?
   \end{itemize}
\end{tcolorbox}

\section{Structure of Awareness}
A little known Soviet psychologist (at least outside Russia) published in the late 60's a book proposing an algebra to mathematically characterize a given reality including not only the specific phenomenon under consideration  but also how the involved persons perceive the whole scenario.
This book appeared translated to English in 1977 with the name \textit{The Structure of Awareness}; the book also appears in a different edition with the name \textit{Conflicting structures} (which is the actual name in Russian).
Although we use a hard-copy from the first version, the second one is more recent and has important parts available on-line \cite{lefebvre2015conflicting}.

In this section, we will present a short review of so-called \textit{algebra of reflexive processes}, which will be of great importance later.
It is worth mentioning that this book is somehow rare and we plan to spread these ideas more and more in non-Russian communities (in Russia, this approach is still active under Dr. Vladimir Lepsky's guidance from the Institute of Philosophy at the Russian Academy of Sciences \cite{reflex}).

\vspace{2ex}
\begin{tcolorbox}[colback=black!5!white,colframe=black!50!black,   colbacktitle=black!75!black,title= Vladimir A. Lefebvre \cite{lef}]
  Born in 1936 in Leningrad, USSR, V. A. Lefebvre is a mathematical psychologist at the University of California, Irvine. He has created equations that are supposed to predict the large-scale consequences of individual actions. Among the parameters in the equations are the self image of the individual and the action as perceived via this self-image. The result is a number expressing the probability that the individual in question will perform a specific action.
\end{tcolorbox} 

\subsection{An Algebra of Reflexive Processes}

The book's first chapter starts defining what reflexive processes are (our highlight) \cite{lefebvre2015conflicting}:
\begin{quote}
\textit{Imagine a room full of crocked mirrors like those in amusement parks. The mirrors are placed at different angles. Let a pencil be dropped. Its fall will be fantastically reflected. Thus, the trajectory, already distorted in the various mirrors, will be further distorted in the avalanche of multiply reflected images. \textbf{A reflexive system can be thought of as a system of mirrors and the multitude of reflexions in them.} Each mirror represents a ``persona'' characterized by a particular position. The vastly complex flow of reflexions represents the reflexive process.}
\end{quote}
A straight differentiation shall be done here: a specific physical process (pencil falling) and the several levels of information processing (reflections, reflections of reflections etc.); the reality of the reflexive system is constituted by these processes.
Therefore, its algebraic representation shall incorporate and differentiate them, accordingly.

Let $T$ be the physical process under consideration. 
This physical process is observed by three information processing elements $X$, $Y$ and $Z$. 
The images of $T$ done by $X$, $Y$ and $Z$ are denoted $Tx$, $Ty$ and $Tz$, respectively.
In other words, $Tx$, $Ty$ and $Tz$ are the reflections of $T$ by $X$, $Y$ and $Z$.
It is important to note that we are not concerning at this stage about how reliable these reflections are but only with their existence.
The reflexive system $\Omega$, whose reality we would like to represent, is $\Omega = T + Tx + Ty +Tz = T (1 +x+y+z)$, where the term $(1 +x+y+z) = \omega$ defines \textit{the structure of awareness}.
Let say that element $Z$ cannot observe $T$ directly, but only through the observation done by $Y$.
The system changed to a new reality $\Omega' = T + Tx + Ty +Tyz = T (1 +x+y+yz)$.

Although both $\Omega$ and $\Omega'$ represent exactly the same physical process $T$, the system has a different structure of awareness ($\omega$ and $\omega'$, respectively), which modifies its existence (i.e. dynamics, development, outcomes etc.).
The proposed notation can capture such -- subtle or even invisible -- structural modification.

The system $\Omega$ may also dynamically change following the physical process $T$ or the information processing procedures $Tx$, $Ty$ and $Tz$, or the system's structure of awareness $\omega$.
Excluding both external interference  and the ``natural'' dynamics of the physical process, the system can be only modified by actions of the elements $X$, $Y$ and $Z$.
Note that not all elements need to take actions; the agents are a subset of information processing elements.
Such actions -- which are always related to a decision rule (random decision is also a rule) -- can modify any part of the system: the physical process itself, the information processing procedures  and the structure of awareness.
If we consider a discrete time $n \in \mathbb{N}$, then we have $\Omega_n$ as a \textit{self-developing reflexive-active system}\footnote{This name is a slightly modification from V. Lepskiy description of the \textit{post-non-classical cybernetics of self-developing reflexive-active environment} \cite{reflex}.}.

A special case of system evolution is related to \textit{the operator of awareness}, denoted by the symbol $*$.
Consider the evolution of the system $\Omega_n$:
\begin{eqnarray*}
&\Omega_0  = & T \\
&\Omega_1  = & \Omega_0 * (1+x) = T * (1+x) = T + Tx\\
&\Omega_2  = & \Omega_1 * (1+y) = (T + Tx) * (1+y) = T + Tx + (T + Tx)y.
\end{eqnarray*}
Putting the sequence in plain words: (n=0) there only exists the physical process $T$; (n=1) element $X$ becomes aware of $T$; (n=2) element $Y$ becomes aware of both $T$ and the information processing $Tx$ that $X$ is doing, i.e. $Y$ constructs its own image of $X$ image of $T$, resulting $Txy$.
Note that the awareness operations represent a dynamical behavior in $n$, but they also lead to a static structure of awareness for the time $n$.
The formalism used to manipulate the algebra of reflexive systems is presented next just as a general reference.

\vspace{2ex}
\begin{tcolorbox}[colback=black!5!white,colframe=black!50!black,   colbacktitle=black!75!black,title= Formalism \cite{lefebvre2015conflicting}]
  \begin{itemize}
   \item Given set of symbols: capital and lowercase letters, and ``$1$''.
   \item Symbols written in sequence without comma are called words.
   \item Words differing in the number of occurrences of ``$1$'' are equivalent; it can always be canceled except when the word consists of a single ``$1$''.
   \item Polynomials are of the form $\omega = \sum_{i=1}^k \alpha_i \; a_i$, where $\alpha_i$'s are Boolean functions and $a_i$ are words. If $a_i = a_j = a$, then  $\alpha_i a_i + \alpha_j a_j =(\alpha_i +\alpha_j)a$.
   \item Addition of polynomials are cumulative: $\omega_1 + \omega_2 = \omega_2 + \omega_1$.
   \item The product of two polynomials $\omega_1 = \sum_{i=1}^k \alpha_i \; a_i$ and $\omega_2 = \sum_{j=1}^l \beta_j \; b_j$ is defined as: $\omega_1 * \omega_2 = \sum_{i=1}^k \sum_{j=1}^l \alpha_i \beta_j \; a_i b_j$.
   \item Postulate: $w^0=1$.
   \item Product is associative, but not generally commutative: $(\omega_1 + \omega_2) * \omega_3 = \omega_1 * \omega_3  + \omega_2 * \omega_3$ and $\omega_3 * (\omega_1 + \omega_2)  = \omega_3 * \omega_1  + \omega_3 * \omega_1$.
  \end{itemize}
\end{tcolorbox} 

\subsection{Poker example}
Let us return to the poker game from the previous section to verify whether  poker is a self-developing reflexive-active system.
Even though poker is defined by its specific rules, it does not exist without players (even if they are all robots).
The matches' dynamics, as indicated before, are determined by a combination of strategic behavior by players and some randomness.
Such a behavior means actions based on processed information from the game reality aiming at specific individual goals.
Therefore, a poker match can be indeed represented by a self-developing reflexive-active system.

The physical process $T$ in this case is determined by the rules (i.e. how the cards are mixed, the betting periods, the order of players etc.).
Considering four players $X$, $Y$, $Z$ and $W$ who have never played together so they do not have any idea about the specific behavior of their opponents (except they want to win, as in any competition).
Before the first match, only the rules exist: $\Omega_0 = T$.
In the first match the system comes to life as $\Omega_1 = T(1 + x + y +z + w)$, i.e. the match is running and the players are aware of the rules of the game, their individual cards and the public information (open cards and bets).
After few matches, only player $X$ inferred how $Y$ behaves based on his perception of $Y$'s aggressive betting behavior.
After this operation of awareness done by $X$, the system becomes: $\Omega_2 = T(1 + x + y +z + w + yx)$. 

As $\Omega_2 \neq \Omega_1$, the system changed so its dynamics and outcomes, regardless of the correctness of the image $Txy$.
From this perspective, the question posed in the end of Section \ref{sec:poker} can be answered.
The awareness operations realized by the players change the structure of awareness of the system.
This is a fully symbolic operation, but it might be somehow controlled (or guided).
Bluffing, for instance, works in this way: a strategic action by one player to create an illusion of the physical reality of the cards, also creating a future image of -- and a certain uncertainty about -- that player (i.e. the others would never be sure about that player attitudes).

Back to our example, let say that player $Y$ purposefully wants to pass the image of aggressive player so other players may create a guided image of him.
We see that $Y$ was indeed able to create such an image in $X$.
This image $Tyx$ modifies how $X$ play in a guided way that might be exploited by $Y$.
And, if $Y$ becomes aware of $X$ image, a new image needs to be included in the game reality: $Tyxy$.
Then: $\Omega_3 = T(1 + x + y +z + w + yx + yxy)$.
And, this could go forever.

\vspace{2ex}
\begin{tcolorbox}[colback=black!5!white,colframe=black!50!black,  colbacktitle=black!75!black,title= Exercise]
  Traffic lights are used to avoid car accidents. The cars stop in the red light, not only because they know that this is what is expected, but also because everyone knows that the others are aware of it and are expecting the same behavior. The traffic light -- acting  as a coordinator -- presupposes that everyone knows how to behave, and that everyone knows that the others know how to behave.
  
  1) Write the structure of awareness equation of a perfectly working traffic light system.
  
  2) Modify the equation so the traffic light system is disturbed. Discuss the situation with a real life example.
\end{tcolorbox} 

\section{The Internet of Things}
The name Internet of Things (IoT) indicates that ``things'' -- in contrast to humans -- also connect to each other, forming a specific network to exchange data.
This is indeed not new; (wireless) sensor networks are around for many decades.
Temperature monitoring in different places provides a simple example.
So, what demarcates the difference between IoT and sensor networks?

The answer is somehow fuzzy, but we identify two interrelated aspects where the differentiation shall take place: scale and structure.
The IoT concept indicates a massive number of (relatively autonomous) sensors building a communication network, whose role is to monitor a specific physical reality to provide information to guide possible actions.
This large number of sensors build a structure of information exchange different from (small-scale) sensor networks.
Such bigger network is what we call IoT. 
In the following box, we reproduce the Wikipedia definition of IoT.

\vspace{2ex}
\begin{tcolorbox}[colback=black!5!white,colframe=black!50!black,  colbacktitle=black!75!black,title= The Internet of Things \cite{IoT}]
The Internet of Things (IoT) is the inter-networking of physical devices, vehicles (also referred to as ``connected devices'' and ``smart devices''), buildings, and other items embedded with electronics, software, sensors, actuators, and network connectivity which enable these objects to collect and exchange data. The IoT allows objects to be sensed or controlled remotely across existing network infrastructure, creating opportunities for more direct integration of the physical world into computer-based systems, and resulting in improved efficiency, accuracy and economic benefit in addition to reduced human intervention. When IoT is augmented with sensors and actuators, the technology becomes an instance of the more general class of cyber-physical systems, which also encompasses technologies such as smart grids, virtual power plants, smart homes, intelligent transportation and smart cities. Each thing is uniquely identifiable through its embedded computing system but is able to interoperate within the existing Internet infrastructure. Experts estimate that the IoT will consist of about 30 billion objects by 2020.
\end{tcolorbox} 

Wikipedia provides a broad definition of IoT.
This definition covers  the previously mentioned sensor networks.
It also includes a new common-sense understanding (e.g. \cite{iot-popular}) that considers IoT as a sensing technology  linked to mobile applications to track individual behaviors related to, for instance, physical activities; the list of possible applications is almost infinite.
While acknowledging all these, we rather use IoT as a very specific concept, opening up a path for a scientific inquiry that shall guide practical deployment.
\textbf{We claim that IoT builds the structure of awareness of specific techno-social cyber-physical systems, which are special cases of self-developing reflexive-active systems.}

\section{Large-scale Infrastructures and IoT}
From Oxford dictionary, infrastructure is \textit{the basic physical and organizational structures and facilities (e.g. buildings, roads, power supplies) needed for the operation of a society or enterprise}.
A more detailed definition is given by the USA National Research Council in its 1987 ``Infrastructure for the 21st Century'' framework for research agenda (see the following box), which includes examples and the social function of large-scale infrastructures.

\vspace{2ex}
\begin{tcolorbox}[colback=black!5!white,colframe=black!50!black,  colbacktitle=black!75!black,title= Large-scale infrastructures \cite{NAP798}]
Public works infrastructure  includes both specific functional modes – highways, streets, roads, and bridges; mass transit; airports and airways; water supply and water resources; wastewater management; solid-waste treatment and disposal; electric power generation and transmission; telecommunications; and hazardous waste management – and the combined system these modal elements comprise. A comprehension of infrastructure spans not only these public works facilities, but also the operating procedures, management practices, and development policies that interact together with societal demand and the physical world to facilitate the transport of people and goods, provision of water for drinking and a variety of other uses, safe disposal of society's waste products, provision of energy where it is needed, and transmission of information within and between communities.
\end{tcolorbox} 

Infrastructures are then the substrate of a specific kind of social relations; as they are not given, but human-built through technology targeting a specific use, they are also technological systems.
Therefore, large-scale infrastructures -- when functioning (i.e. when things are flowing) -- are techno-social systems, 
IoT clearly fits in this definition.
We are, however, more interested in the symbolic function of IoT as a layer of information processing and distribution (although we could, and should, analyze its actual physical infrastructure).
IoT builds symbolic -- cybernetic -- reflection of the physical infrastructure to guide actions to achieve a prescribed goal (e.g. improve the flow of cars in the peak hour).
\textbf{IoT is the structure of awareness of techno-social cyber-physical systems.}

\vspace{2ex}
\begin{tcolorbox}[colback=black!5!white,colframe=black!50!black,  colbacktitle=black!75!black,title= Techno-social cyber-physical  systems \cite{vespignani2009predicting,CPS}]
Modern techno-social systems consist of large-scale physical infrastructures (such as transportation systems and power distribution grids) embedded in a dense web of communication and computing  infrastructures whose dynamics and evolution are defined and driven by human behavior.
  \tcblower
A cyber-physical system (CPS) is a mechanism controlled or monitored by computer-based algorithms, tightly integrated with the internet and its users. In cyber-physical systems, physical and software components are deeply intertwined, each operating on different spatial and temporal scales, exhibiting multiple and distinct behavioral modalities, and interacting with each other in a myriad of ways that change with context.
\end{tcolorbox}

\section{Example: Electricity Power Grids}
We now know that IoT  is more than a simple mobile phone application to measure physical activities or a sensor network that monitors temperature.
The question  is how this structure of awareness would change the dynamics of large-scale infrastructures.
The first step to answer this question is to define the (techno-social cyber-physical) system to be analyzed. 
Then, we need to select a specific phenomenon (or a set of phenomena) to be investigated.
Only after these two steps, the method of research and the interventions using IoT can be evaluated.
In this section, we will follow our previous research \cite{nardelli2017smart}, but here we carefully describe our methodology.

Although this procedure seems straightforward, the actual practice involves several difficulties to be discussed later.
We exemplify our analysis by looking at the effects of the so-called ``smart appliances'' when used to help stabilizing electricity power grids.

\subsection{Defining the system}
How to precisely define the electricity power grid?
As any large-scale infrastructure, such a system is related with flows.
As the name suggest, it relates to transferences of energy in form of electricity that is needed by households, industries, hospitals etc. (i.e. a social need).
Remembering that power is defined as flow of energy, electricity power grids can be then defined as a system by the following operation: interchange (transport) of electricity.

\vspace{2ex}
\begin{tcolorbox}[colback=black!5!white,colframe=black!50!black,  colbacktitle=black!75!black,title= Electricity power grid as a system]
A given \textbf{system} defines itself by its peculiar operation; in large-scale infrastructures this operation is related to flows.
To sustain the system, two things are necessary: (1) production and demand of whatever is flowing and (2) reproduction of the conditions allowing the flows.
While these two aspects are internal and necessary to keep the system alive, they are structurally dependent on other external factors as, for example, laws, habits, investments and/or weather conditions.
\tcblower
\textbf{Electricity power grid as a system} is defined by the operation \textit{electric power interchange}. 
Condition (1): It is produced by power plants or distributed generation while consumed by households, industries etc.
Condition (2): Power lines need to be maintained, operational decisions to ensure power quality is needed, distribution decisions (regulated by market or planing) etc.
Structural (external) relations: Investments in production and power lines, training of specialized personal, raw material for power plants, regulatory laws, weather allowing for wind generation and many other aspects. 
\end{tcolorbox} 

\subsection{Defining the phenomenon and interventions}
Given the system, its condition of existence as such and the structural relations involved, we need to narrow down our analysis to a specific phenomenon (or set of phenomena) that we want to better understand or intervene.
As an illustration we select the frequency stability that reflects the real-time balance of supply and demand in the AC connected grid, where every connected element experiences an alternate (sinusoidal) electric current/voltage with the same frequency.
In very simple terms: if there is more power supplied than demanded, then frequency assumes higher values, and vice-versa.
This effect is physical so it happens in the physical layer; however, the frequency (if to be used in any way), can be measured or sensed by some specific devices.
The measured frequency is then part of the information layer and reflects the physical reality.
Only through this lens, agents can take (informed) actions targeting at a prescribed objective.
Our focus here is to analyze possible interventions focusing on condition (2).
The goal is to guarantee the power quality of the transmission, maintaining the frequency under a preset fixed value with a small variation.
Following the European standard, frequency shall be in the range $50 \pm 0.1$ Hz. 

As frequency reflects the balance between supply and demand, we can use reactive (``smart'') appliances that work in cycles (e.g. fridges) to advance or postpone their operation to help balancing the power grid.
The appliances are then agents in the regulatory layer that (re)act to the information about the physical system by modifying (or not) their individual behavior.

\vspace{2ex}
\begin{tcolorbox}[colback=black!5!white,colframe=black!50!black,  colbacktitle=black!75!black, title= Think about]
\begin{itemize}
\item How would you design an appliance to achieve this goal?
\item What would you imagine the system dynamics with these appliances?
\item Does the number of reactive appliances affect the system stability? 
\item Is there a structure of awareness in the system?
\end{itemize}

\end{tcolorbox} 

\subsection{Toy-model}
To answer scientifically those questions, one needs to model the system so that the phenomenon under analysis can be captured.
The scientific research needs to identify the essential features that shall be captured and the others that can be ignored (at least in the first stages).
In this case, we decided to study the effects of the reactive appliances by proposing a \textit{toy-model} instead of using a high realistic power system simulator.
We argue that the proposed toy-model is the simplest way to emulate the system phenomenon under investigation while considering other aspects as perfect, given or just negligible.
Be no mistaken here: the results obtained therefrom need to be put in their place.
They are not the reality and do not imply that this is what is going to happen; it shows a possible qualitative behavior that is scientifically grounded in its own narrow, abstract, domain.
Any extrapolation of the results obtained through the toy-model to real-world conditions must be carefully done case-by-case. 

\begin{figure}
	\centering
	\includegraphics[width=0.75\columnwidth]{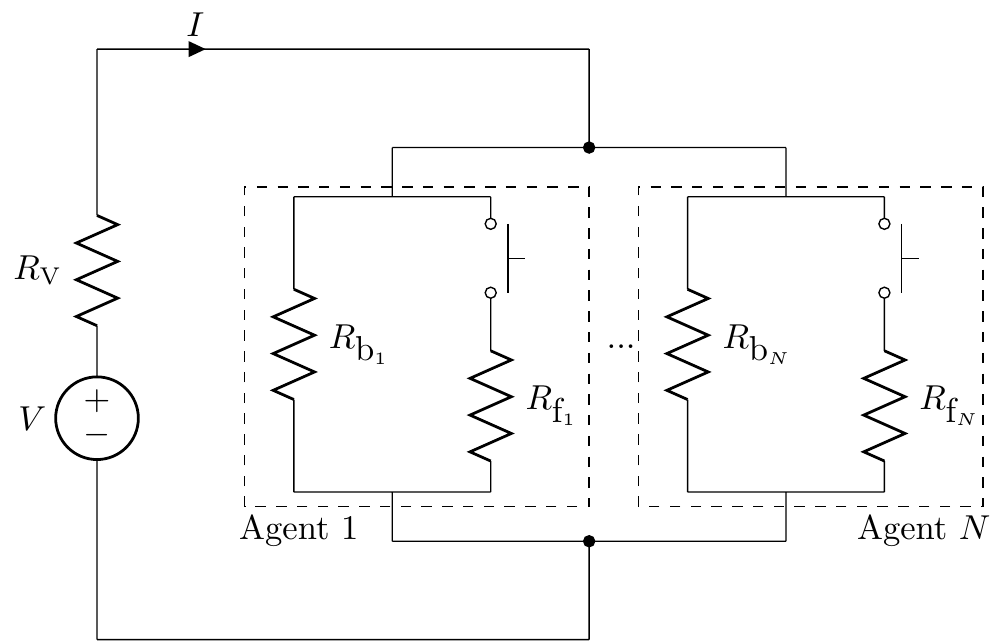}
	\caption{Electrical circuit representing the physical layer of the system. The circuit is composed by a power source $V$ and its associate resistor $R_\textup{V}$, and resistors of in parallel, generating a current $I$. These resistors are related to $N$ agents, where each agent $i=1,...,N$ has a basic load $R_{\textup{b}_i}$ and controls a flexible load $R_{\textup{f}_i}$.}
	\label{fig:System} 
\end{figure}

For our power grid example, we model the AC power grid by a simple DC electric circuit as presented in Fig. \ref{fig:System}.
This is the physical layer of the system, where the voltage experienced by the agents emulates the frequency in the power grid considering a controlled supply.
Agents are set to turn on and off their loads in cycles that may be postponed or advanced depending on the voltage experience by them.
For example, if the supply is decreased by some amount, reactive agents postpone the cycle of adding such a flexible load (resistor).

\subsection{Case study}
\begin{figure}
	\centering
	\includegraphics[width=\columnwidth]{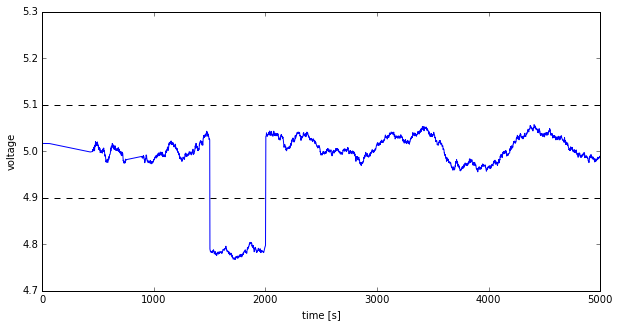}
	\caption{Controlled input. A drop in voltage below the threshold so a reaction in reactive agents can be triggered.}
	\label{fig:in} 
\end{figure}

\begin{figure}
	\centering
	\includegraphics[width=\columnwidth]{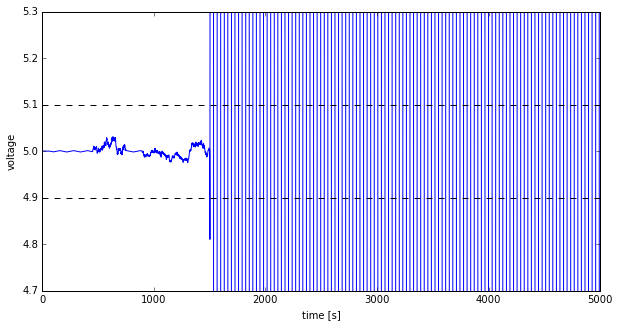}
	\caption{Case study 1. All agents react to the signal. Collective behavior determined by the system's structure of awareness and the agents' respective reactions.}
	\label{fig:out1} 
\end{figure}

Let us consider the following case.
All agents are reactive to the voltage.
The system structure of awareness is stable and characterized as $\Omega = T (1 + \sum a_i)$, where $a_i$ relates to agent $i$.
In this first case, all agents sense the same original signal from the physical layer (``they see the same physical reality'')  and react to the controlled signal in the same way (we can say that the agents are homogeneous).
Fig. \ref{fig:in} shows an example of the voltage experienced by the agents.
To illustrate the system dynamics considering interactions between the physical circuit, the structure of awareness and the agents' decision rules, we emulate a voltage drop during a given period.
An example of output is presented in Fig. \ref{fig:out1}.

At the first sight, this is a remarkable -- unexpected and undesired -- behavior.
\textbf{How is it possible that the system designed to have (non-competitive) agents with the same objective (stabilizing the voltage within a given range) lead to the opposite outcome, worsening the situation they were designed to solve?}
The answer is quite simple as soon as we consider the three-layer model: the structure of awareness of the system implies that the agents only look at the information about the physical system without considering other agents (i.e. \textit{selfish} decisions, even without competition).
In this case, the physical signal trigged all agents to react in the same way.
When the signal dropped, all agents postpone their cycles, synchronously. 
Then, we see a voltage overshoot due to the collective behavior.
This overshoot led to a voltage increase above the threshold, triggering a new synchronization (but now of agents advancing their cycles).
Consequently, we see a new voltage drop and the system dynamics become unstable and internally impossible to be stabilized, always out of the normative limits.

\vspace{2ex}
\begin{tcolorbox}[colback=black!5!white,colframe=black!50!black,  colbacktitle=black!75!black, title= To think about]
\begin{itemize}
\item How to solve this instabilities?
\item What kind of structure of awareness could change this behavior?
\item What kind of decision rules could change this behavior?
\end{itemize}
\end{tcolorbox} 

\subsection{Possible solutions}
To solve this problem, the collective behavior emerging from the specific combination of structure of awareness and agents' reaction to the information signal (frequency) needs to be controlled.
For example, a new entity can appear in the system as the central controller $C$ of the flexible loads.
In this case, the structure of awareness become: $\Omega = T (1 + c +  \sum a_i) + Tc(\sum a_i)$, where $Tc(\sum a_i)$ is the information received by the agents $A_i$ about how the control element $C$ sees reality $T$ and the expected reactions from the different agents.
Note that now there are two kinds of $Tca_i$'s: (1) React to the signal $Ta_i$, or (2) Do not react to the signal $Ta_i$.
The central controller need them to compute how many agents are necessary to react, who must act, and inform them accordingly.

\begin{figure}
	\centering
	\includegraphics[width=\columnwidth]{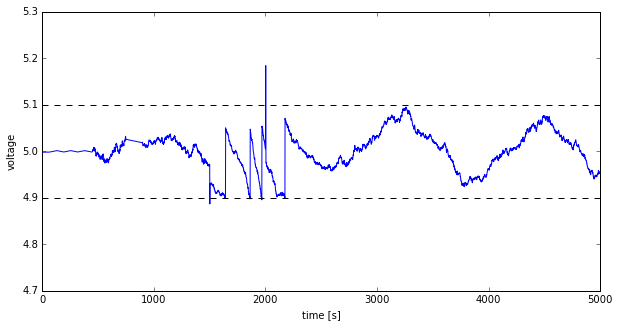}
	\caption{Possible solution to case study 1. Each agent reacts to the signal with a given probability $p$. Collective behavior is controlled. This solution is only feasible if every agent knows that the other are behaving in similar away.}
	\label{fig:out2} 
\end{figure}

A similar idea can be implement without centralized control.
The agents are then designed to have a randomized decision so that, when an event that needs a reaction happens, each individual reaction occurs with a given probability $p$.
But, this only works if all agents know that every other agent in the system will act in the same way.
Hence, we have the following structure of awareness: $\Omega = T(1+ \sum_i a_i + \sum_j\left( \sum a_i\right)a_j)$.
This means that every agent $A_i$ has an image of how every other agent (including itself) will act when experience the input signal.
The output using this strategy can be seen in Fig. \ref{fig:out2}, where one see that the effects presented in Fig. \ref{fig:out1} were mitigated.
Note that this structure of awareness is static and built by design, not guided by the system dynamics.

\vspace{2ex}
\begin{tcolorbox}[colback=black!5!white,colframe=black!50!black,  colbacktitle=black!75!black, title= Discussions and simulations]
\begin{itemize}
\item Problems related to this solution.
\item Discussing about possible other solutions and their problems.
\item Simulate other proposed scenarios, available at \cite{simulation}.
\item Describe their decision processes in relation to the structure of awareness.
\end{itemize}
\end{tcolorbox} 

\section{Final remarks}
In this text, we propose to go beyond the narrow perspective that the Internet of Things are merely a new mask of the old sensor networks, or a combination of simple devices connected to mobile phone applications.
We argue that the IoT provides the structure of awareness of large-scale infrastructures, which are a specific class of self-developing active-reflexive systems called cyber-physical techno-social systems.
We illustrated how the structure of awareness -- normally invisible at the first sight -- actually modifies the dynamic of systems.
In this case, a model needs to include physical, information and regulatory interacting layers so that interventions could be properly designed.

We analyzed a specific case of frequency control in electricity power grids through a toy model.
For this case, we analyzed a problematic case as well as a possible solution to it.
We also proposed an open-source simulation to test other interventions.
\textbf{All in all, we expect to put light on the theoretical and practical importance of a multi-layer model of systems, including explicitly its often implicit structure of awareness.}
More information about this line of research can be found in \cite{pedro-homepage}.

\vspace{2ex}
\begin{tcolorbox}[colback=black!5!white,colframe=black!50!black,  colbacktitle=black!75!black, title=To think about]
There are many price-based or market-based solutions to the power grid.
One is the demand-response where the consumer would follow the supply informed by the electricity price (lack of supply for a given demand implies higher prices and vice-versa).
Price can be seen as a universal signal to every agent.
Users are then expect to do both: maximize their own utility (selfishness cost minimization) and balance the system (reduce or increase demand, following supply).
Think about this issue using the ideas developed here, posing possible problems and interventions based on IoT.
\end{tcolorbox}

\bibliographystyle{IEEEtran}
%\bibliography{ref}

\end{document}